\documentclass[aps,prl,floatfix,reprint,amsmath,amssymb]{revtex4-1}

\usepackage{bm}
\usepackage[pdftex]{graphicx}

\newcommand{\vect}[1]{\mathbf{#1}}
\newcommand{\hb}{\mbox{H-bond}}
\newcommand{\wb}{\ensuremath{W_{\mathrm{break}}}}

\begin{document}

\title{Atomistic Hydrodynamics and the Dynamical Hydrophobic Effect in Porous Graphene}

\author{Steven E. Strong}
\author{Joel D. Eaves}
\email{Joel.Eaves@colorado.edu}
\affiliation{Department of Chemistry and Biochemistry, University of Colorado, Boulder, Colorado, United States}
\date{\today}

\begin{abstract}
Mirroring their role in electrical and optical physics, two-dimensional crystals are emerging as novel platforms for fluid separations and water desalination, which are hydrodynamic processes that occur in nanoscale environments. For numerical simulation to play a predictive and descriptive role, one must have theoretically sound methods that span orders of magnitude in physical scales, from the atomistic motions of particles inside the channels to the large-scale hydrodynamic gradients that drive  transport. Here, we use constraint dynamics to derive a nonequilibrium molecular dynamics method for simulating steady-state mass flow of a fluid moving through the nanoscopic spaces of a porous solid. After validating our method on a model system, we use it to study the hydrophobic effect of water moving through pores of electrically doped single-layer graphene. The trend in permeability that we calculate does not follow the hydrophobicity of the membrane, but is instead governed by a crossover between two competing molecular transport mechanisms.
\end{abstract}
\maketitle

Numerical techniques, rooted in theory, are indispensable tools in the study of liquids and fluids. On microscopic length and time scales, statistical mechanics underpins the molecular dynamics~(MD) methods for systems at thermal equilibrium~\cite{frenkel_understanding_2001}. On macroscopic scales, continuum hydrodynamics can describe fluids driven away from equilibrium~\cite{bird_transport_2006}. But it remains unclear how one should simulate an atomistic system away from equilibrium~\cite{zhu_pressure-induced_2002,huang_method_2011,frentrup_transport_2012,takaba_molecular_2007,huang_molecular_2006,muller-plathe_simple_1997}. This gap in knowledge makes it difficult to model processes on the mesoscale, such as water desalination, gas separation, and cellular transport. In these systems, gradients in continuous fields, like density and pressure, drive flow through bottlenecks that admit only a few particles at a time~\cite{hummer_water_2001,kalra_osmotic_2003,corry_designing_2008,suk_water_2010,cohen-tanugi_water_2012,wang_water_2012,konatham_simulation_2013,xu_nonequilibrium_1999,vieira-linhares_non-equilibrium_2003,skoulidas_molecular_2004,zhu_pressure-induced_2002,koenig_selective_2012,schrier_helium_2010,nair_unimpeded_2012}. These processes require computational models to be theoretically rigorous and accurate across orders of magnitude in physical scales.

Hydrodynamic approaches are rooted in continuum models that inherently break down on atomic scales~\cite{bird_transport_2006}. Conversely, microscopic MD simulations only generate rigorously accurate dynamics for closed and isolated systems~\cite{nose_molecular_1984}. These systems can be coupled to a heat bath to generate static averages consistent with the canonical ensemble, but the thermostats that do this are not unique.~\cite{frenkel_understanding_2001}.  The dynamics generated under various thermostatting schemes can be quite different, even at thermal equilibrium~\cite{nose_molecular_1984,frenkel_understanding_2001}.
In nonequilibrium MD simulations, both an external force and a thermostat counterbalance to maintain steady state~\cite{evans_non-newtonian_1984,evans_statistical_2008}. The implementation of these two components is likewise not unique.  Away from equilibrium, the interaction between the thermostat and an external driving force can produce results that are manifestly unphysical~\cite{evans_shear_1986,harvey_flying_1998,bernardi_thermostating_2010,heyes_nonequilibrium_1985}.

\begin{figure*}
\includegraphics[width=\linewidth]{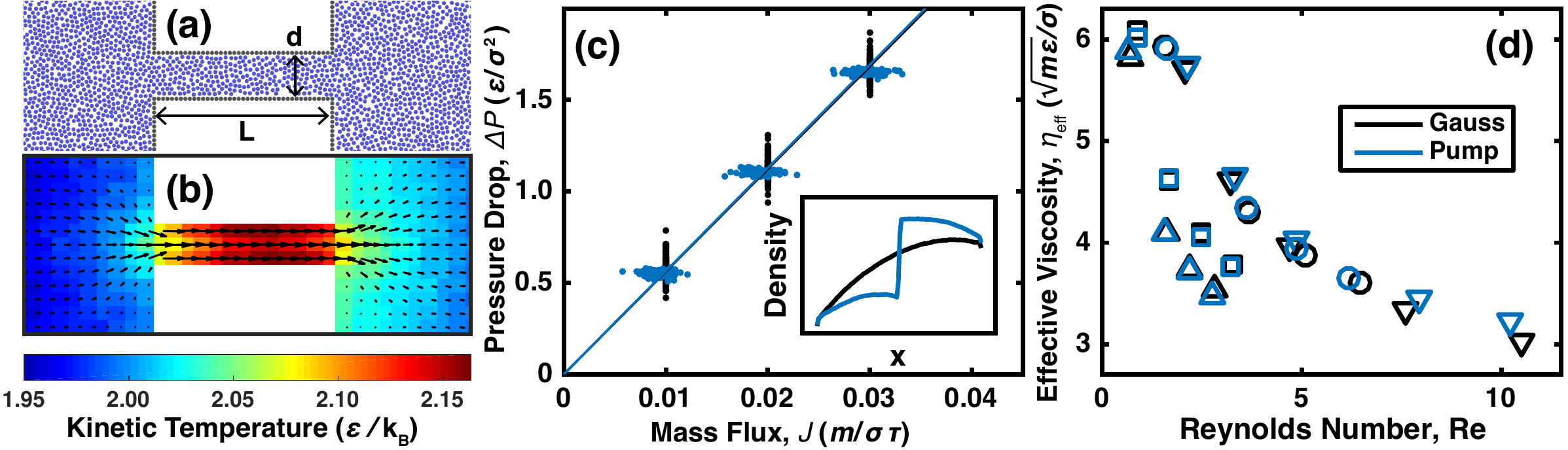}
\caption{(a) Closeup of a snapshot from a 2D Lennard-Jones simulation evolving under GD, including variables for the length~($L$) and diameter~($d$) of the channel. (b) Steady-state kinetic temperature~(color) and velocity field~(vectors), $\vect u(\vect r)$, averaged over time at $\mathrm{Re}=3$. Hydrodynamic variables, like $\vect u(\vect r)$, and associated gradients in density, temperature, and pressure develop naturally under the imposed constraints. (c) The pressure drop as a function of the mass flux, $J$, for both GD and the pump method in 2D Lennard-Jones simulations at various flow rates, with 96 trials at each flow rate. The slope of these data determines the effective viscosity, $\eta_\mathrm{eff}$~(see Eq.~\ref{visc} and the SI). Panel~(d) compares $\eta_\mathrm{eff}$ for the two methods at various $L$ and $d$, plotted as a function of Re at maximum $J$. The symbol shape indicates the diameter~($d$) of the channel: $\boldsymbol\bigtriangleup$~($d=18~\sigma$), $\boldsymbol\square$~($d=14~\sigma$), $\boldsymbol\circ$~($d=8~\sigma$), $\boldsymbol\bigtriangledown$~($d=4~\sigma$). The computed $\eta_\mathrm{eff}$ of the two methods match well at low Re but show increasing differences for narrow channels~($\boldsymbol\bigtriangledown$) as Re increases. The inset to panel~(c) shows the density profile along the direction of flow,~$x$. The inset is a periodically wrapped image of~(a) and~(b), with the pore and its periodic image at the left and right edges of the inset. The density is discontinuous in the pump region of the pump method, but smooth in GD. The changes in density shown here are $\pm$ 3\% from the average bulk density~(see the SI).\label{fig1}}
\end{figure*}

In this manuscript, we develop a method for simulating atomistic systems in nonequilibrium steady states of mass flow.  Our method, which we call Gaussian dynamics~(GD), finds the equations of motion that are consistent with a minimal set of constraints, much like early system--bath coupling schemes devised in MD methods~\cite{evans_statistical_2008}. We constrain only the total mass current and kinetic temperature. Gradients in hydrodynamic fields such as velocity, density, temperature, and pressure arise naturally~(Fig.~\ref{fig1}b). 
We test GD using a simple two-dimensional~(2D) liquid flowing through channels of various geometries at various Reynolds numbers~(Re).

We then use GD to study the permeability of porous 2D crystals. Porous 2D crystals offer a new paradigm of atomically thin semipermeable membranes for gas and liquid separations~\cite{koenig_selective_2012,schrier_helium_2010,nair_unimpeded_2012,drahushuk_analysis_2016,jain_heterogeneous_2015}, and have important applications in water desalination through reverse osmosis~\cite{surwade_water_2015,kalra_osmotic_2003,corry_designing_2008,suk_water_2010,wang_water_2012,cohen-tanugi_water_2012,cohen-tanugi_water_2014,han_ultrathin_2013,konatham_simulation_2013,heiranian_water_2015,jain_heterogeneous_2015}. At low Re, where most reverse osmosis devices operate~\cite{dashtpour_energy_2012}, one expects the system to be close enough to equilibrium that linear response theory is accurate~\cite{zhu_collective_2004}. But for porous 2D crystals to function as high-fidelity separators, the pores must be so small that only a few water molecules occupy them at a time.  Therefore, as we show here, permeabilities computed from linear response average poorly, and it is more practical to perform nonequilibrium simulations. 

We study the dynamical hydrophobic effect in porous 2D crystals using electrically doped graphene, which has a continuously tunable hydrophobicity~\cite{ostrowski_tunable_2014} and is experimentally realizable~\cite{surwade_water_2015,farmer_chemical_2009,bao_situ_2012}. 
In water desalination applications, the hydrophobic effect can play two counteracting roles: Water facing a hydrophobic sheet could feel a penalty toward wetting the pore, thereby lowering the permeability~\cite{lum_pathway_1997,lum_hydrophobicity_1999,bolhuis_transition_2000,brovchenko_water_2004}. But water might also adhere less strongly to a hydrophobic surface, which would lower the friction and increase the permeability~\cite{willard_molecular_2014}. This latter effect is purely dynamical in nature. We find that the observed behavior is more complex than either of these scenarios would predict.  Understanding it requires a detailed picture of the microscopic transport mechanism, which we build using a Markov model.

We begin by deriving GD.\@ For a system of~$N$ fluid particles with masses~$\{m\}$ at positions~$\{\vect{r}\}$, suppressing notation for all implicit time dependence, the flow constraint is
\begin{equation}\label{flow}
\vect{g}_\mathrm{f} \big(\{\vect{r},\vect{\dot r}\}\big)=\sum_{i=1}^N m_i\big(\vect{\dot r}_i - \vect{u}(\vect r_i)\big) = 0, 
\end{equation}
where $\vect{u}(\vect r)$ is the streaming velocity of the fluid evaluated at point ${\vect r}$, and dots denote time derivatives. This constraint is nonholonomic and cannot be treated easily using Euler--Lagrange or Hamiltonian dynamics~\cite{cronstrom_nonholonomic_2009,flannery_elusive_2011}, so we turn instead to Gauss's principle of least constraint~\cite{gauss_uber_1829,evans_nonequilibrium_1983,evans_statistical_2008}. The Gaussian cost function $C$ is
\begin{equation}\label{cost}
C\big(\{\vect{ \ddot r}\}\big)=\frac{1}{2}\sum_{i=1}^{N}m_i {\left( \vect{\ddot r}_i - \frac{\vect{F}_i}{m_i}\right)}^{2} +   \lambda_\mathrm{EM} G_\mathrm{EM} + \boldsymbol \lambda_\mathrm{f} \bm\cdot \frac{d\vect g_\mathrm{f}}  {dt},
\end{equation}
where $\vect F_i=-\boldsymbol\nabla_i U$ is the force on particle~$i$ from the potential function $U$, $\lambda_\mathrm{EM}$ and $\boldsymbol\lambda_\mathrm{f}$ are Gaussian multipliers, and $G_\mathrm{EM}$ is the Evans-Morriss constraint on the kinetic temperature and molecular geometries~\cite{hoover_high-strain-rate_1982,evans_computer_1983,evans_nonequilibrium_1983,evans_statistical_2008,edberg_constrained_1986,morriss_constraint_1991}. The accelerations that minimize this cost function satisfy the equation of motion
\begin{equation}\label{eq:eom}
m_i \vect{ \ddot r}_i = \vect{F}_i + {\vect f}_i- m_i\xi\big( \vect{\dot r}_i - {\vect u}({\vect r}_i) \big) - m_i {\vect I},
\end{equation}
where ${\vect f}_i$ is the rigid bond constraint on particle~$i$ and $\xi$ is the drag coefficient of a profile-unbiased isokinetic thermostat~\cite{hoover_high-strain-rate_1982,evans_computer_1983,evans_nonequilibrium_1983,evans_statistical_2008,edberg_constrained_1986,morriss_constraint_1991,evans_shear_1986}. A comprehensive derivation of Eq.~\ref{eq:eom} appears in the Supporting Information (SI). The flow constraint, Eq.~\ref{flow}, introduces an external force,~${m_i\vect I}$, into Eq.~\ref{eq:eom}.
\begin{equation}
{\vect I}=\frac{1}{M}\sum_{j=1}^N\vect{F}_j,
\end{equation}
where $M=\sum_{i=1}^N m_i$ is the total mass of the fluid. ${\vect I}$ is weak, fluctuating in time, and uniform in space (see the SI). It acts as a fluctuating gravitational field that maintains the mass current, counteracting the virtual work required to hold a set of wall or membrane atoms fixed in space. Computing $\vect I$ scales as ${\cal O}(N)$, so it adds little computational burden.  As in the isokinetic thermostat, one can solve for $\xi$ iteratively at each time step. Instead, we take the more computationally efficient approach and fix the average kinetic temperature using a profile-unbiased Nos\'e--Hoover thermostat~\cite{nose_unified_1984,hoover_canonical_1985,evans_shear_1986,evans_statistical_2008}. 

Equation~\ref{eq:eom} is a central result of this manuscript. While simple, it is theoretically rooted in constraint dynamics and stands in contrast to  \emph{ad-hoc} approaches that employ some mixture of external forces, particle swaps, and thermostats~\cite{zhu_pressure-induced_2002,huang_molecular_2006,huang_method_2011,muller-plathe_simple_1997,kannam_slip_2012,frentrup_transport_2012,takaba_molecular_2007}. In the context of nonequilibrium statistical mechanics, GD, a constant current protocol, is a Norton ensemble method. In this manuscript, we compare results from GD with its conjugate Th\'evenin ensemble, or fixed gradient method, Zhu, Tajkhorshid, and Schulten's ``pump method''~\cite{zhu_pressure-induced_2002}. Where possible, we also compare to the equilibrium predictions from linear response theory~\cite{zhu_collective_2004}. 


A simple 2D Lennard-Jones fluid flowing through a channel provides a computationally feasible test system~(Fig.~\ref{fig1}a). To compare the GD and pump methods, we draw on the Hagen-Poiseuille~(HP) law from hydrodynamics to calculate an effective viscosity,~$\eta_{\mathrm{eff}}$,
\begin{equation}\label{visc}
\eta_{\mathrm{eff}} = \frac{d^2\rho\Delta P}{12LJ},
\end{equation}
 which relates the mass flux, $J=\rho_\mathrm{t} |\vect u|$~(see the SI), to the pressure drop,~$\Delta P$, applied across a channel of length~$L$ and diameter~$d$~(Fig.~\ref{fig1}a). Note the distinction between $\rho$, the mass density of the bulk fluid, and $\rho_\mathrm{t}$, the total mass density of the fluid over the \emph{entire} simulation box, including volume excluded by the channel walls~(see the SI). We certainly do not expect the HP law to be quantitative on these length scales, but merely use it as a practical means to discuss the relationship between the current and the pressure drop for channels of various geometries in a consistent way~(Fig.~\ref{fig1}c). The Norton and Th\'evenin ensembles should give similar results for the effective viscosity~$\eta_{\mathrm{eff}}$, regardless of the fundamental inaccuracy of the HP law~(see the SI).  

We compute the pressure as a function of position in the simulations using the zeroth-order Irving--Kirkwood approximation, which we apply only where it is valid, away from the channel walls~\cite{irving_statistical_1950,todd_pressure_1995}.  This method is convenient and accurate, but not unique~\cite{hafskjold_microscopic_2002,heinz_calculation_2007,zimmerman_calculation_2004,irving_statistical_1950,todd_pressure_1995}. The pressure drop, $\Delta P$, comes from a linear extrapolation to the edges of the channel. In GD, the fluctuating acceleration, $\vect I$, adds a hydrostatic pressure, which we simply subtract before computing $\Delta P$~(see the SI). This technique differs from others reported in the literature~\cite{zhu_pressure-induced_2002,huang_method_2011}. 

We compare GD and the pump method over a range of Re,
\begin{equation}
  \mathrm{Re} =\frac{\overline{u_{\mathrm{in}}}L\rho}{\eta},
\end{equation}
where $\overline{u_{\mathrm{in}}}$ is the time-averaged center-of-mass velocity of the fluid inside the pore and $\eta$ is the bulk viscosity of the fluid~(see the SI). The effective viscosity~(Eq.~\ref{visc}) computed using GD compares well with that computed from the pump method, particularly at low Re~(Fig.~\ref{fig1}d).  At larger Re~($\mathrm{Re}>5$) there is more disagreement. It would be informative to simulate higher Re and narrower channels, but these regimes take a prohibitive amount of computational time to reach steady state~(see the SI). Some, but not all, of the disagreement at higher Re is due to the thermostat conventionally used in the pump method, which is not Galilean-invariant~\cite{zhu_pressure-induced_2002,huang_method_2011,frentrup_transport_2012}. To correct for this, we have amended the original pump method to include a profile-unbiased thermostat. This increases the agreement between the two methods at higher Re, but it does not fully account for the discrepancies observed~(see the SI). 

For a semipermeable membrane, an important figure of merit is the permeability, $p$, 
\begin{equation}\label{eq:perm}
p=k_B T \frac{q}{\Delta P},
\end{equation}
where $q$ is the flow rate~(molecules/time), which is proportional to the mass flux, $J$~(see the SI). The permeability is inversely related to the effective viscosity~(Eq.~\ref{visc}).

We compute the permeability of porous single-layer graphene over a range of voltages applied to the sheet using GD, the pump method~\cite{zhu_pressure-induced_2002}, and linear response theory~\cite{zhu_collective_2004}. For GD and the pump method, we compute the permeability using Eq.~\ref{eq:perm} with the slope of $q$ versus $\Delta P$~(analogous to Fig.~\ref{fig1}c). At equilibrium, we compute the permeability using linear response theory, as described in Ref.~\citenum{zhu_collective_2004}. As in Ref.~\citenum{ostrowski_tunable_2014}, we use the rigid SPC/E water model~\cite{berendsen_missing_1987,ryckaert_numerical_1977,andersen_rattle:_1983}, a standard potential for the carbon--oxygen interaction~\cite{werder_watercarbon_2003}, and find the effective charge per carbon atom as a function of voltage from the dispersion relationship of graphene~\cite{ostrowski_tunable_2014}. All carbon atoms have the same partial charge, with the charge placed at the atomic centers of each carbon. The SI contains the simulation details.

\begin{figure}
\includegraphics{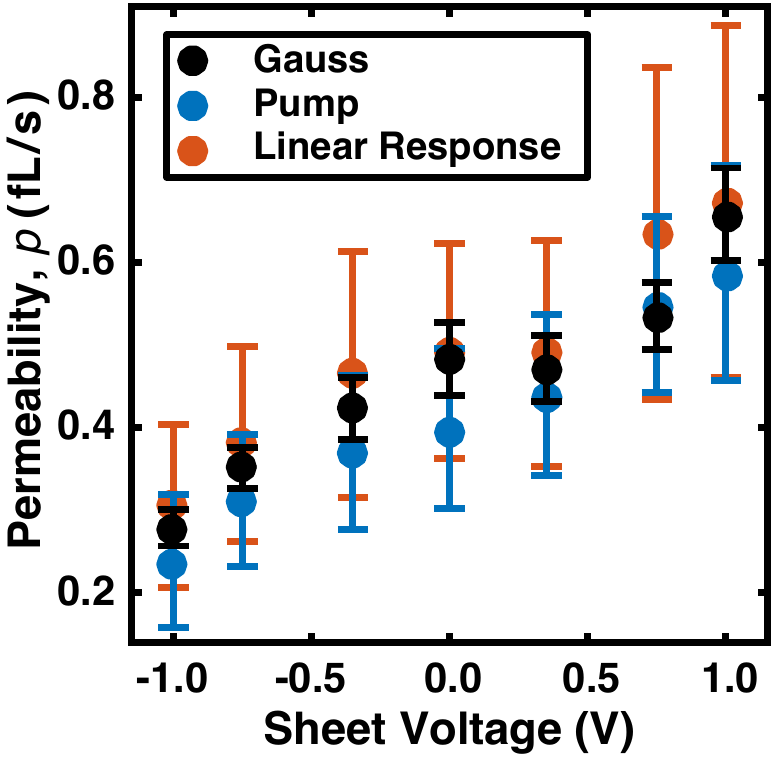}
\caption{Permeability in femtoliters/second (Eq. \ref{eq:perm}) of a single pore in a graphene sheet as a function of the voltage applied to the sheet, reported in volts. The hydrophobicity of the graphene sheet, calculated in Ref.~\citenum{ostrowski_tunable_2014}, does not follow the permeability shown here.\label{fig3}}
\end{figure}

Using contact angle measurements, MD simulations of water droplets on graphene have shown that graphene becomes more hydrophilic at both positive and negative applied voltages~\cite{ostrowski_tunable_2014}. In light of these simulations, our results show that the hydrophobicity of the sheet does not predict the permeability~(Fig.~\ref{fig3}). The permeability of the sheet is higher at positive voltages~(excess electrons) but lower at negative voltages~(excess holes), even though the sheet is more hydrophilic in both regimes~\cite{ostrowski_tunable_2014}. The size of the error bars illustrates the difficulty of converging these calculations, and it is only with GD that a statistically significant trend appears. For similar computational costs and for all simulations and quantities reported here, the standard errors are smaller for GD than those for either of the other methods~(see the SI). 

\begin{figure*}
\includegraphics[width=\linewidth]{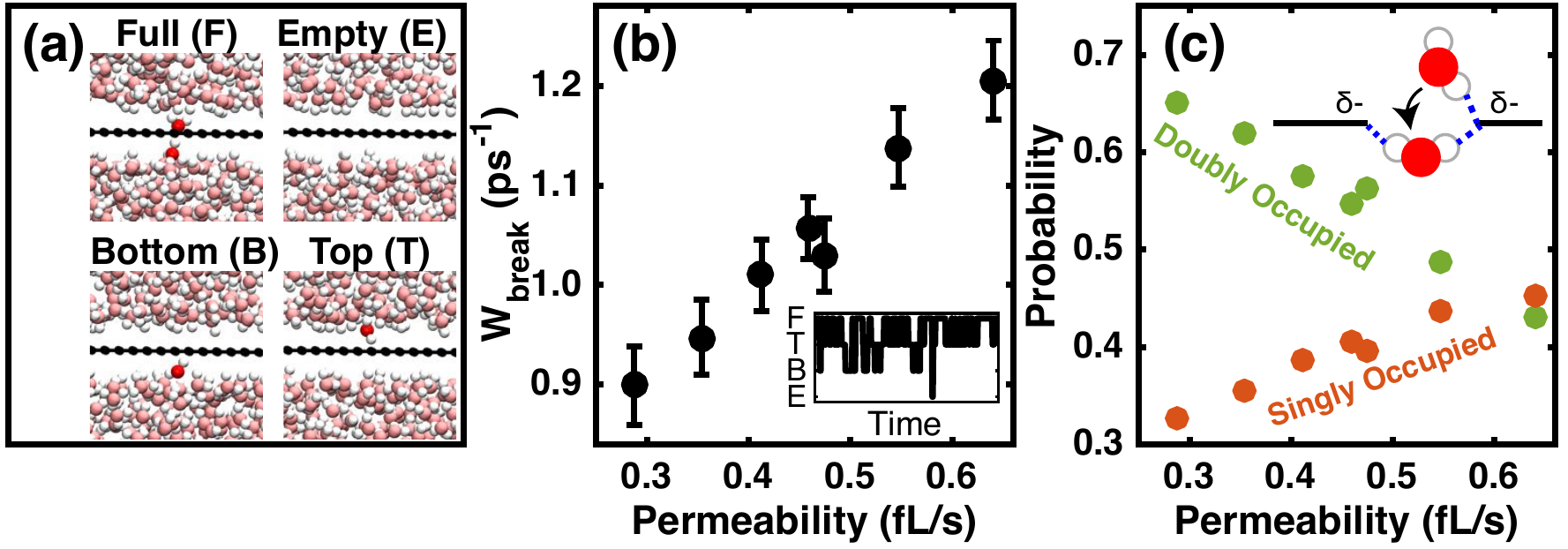}
\caption{(a) The four states used in the microscopic Markov model for transport. The inset to panel~(b) shows 20~ps of a time series of this Markov process, from which we compute the transition probabilities~(see the SI). Panel~(b) shows \wb, the rate at which a pair of molecules in the pore break their hydrogen-bond~(H-bond), which we interpret as a proxy for the evaporation--condensation transport rate~(see text). Panel~(c) shows the steady-state probabilities of singly~(orange) and doubly~(green) occupied states. Both \wb\ and the probability of a singly occupied state are correlated with the permeability, while the doubly occupied state is anticorrelated with it. The inset to panel~(c) illustrates how the limber hydrogen atoms of a water molecule form contacts with a negatively charged sheet, enabling H-bond dissociation and single occupancy, and encouraging the evaporation--condensation mechanism.\label{fig4}}
\end{figure*}

The discrepancy between the permeability and the hydrophobicity suggests that passage dynamics are not dominated by a large-scale collective hydrophobic effect, like capillary wetting~\cite{willard_molecular_2014}.  We instead suspect that microscopic motions control the transport dynamics in pores with dimensions comparable to a water molecule. To test this hypothesis, we coarse-grain the occupancy of the channel and develop a stochastic Markov model of the transport process. The pore is small enough that passage is single-file~(see the SI), so there are only four Markov states, depicted in Fig.~\ref{fig4}a. We run simulations at equilibrium and compute the transition probabilities and steady states of the Markov process directly from the time series~(inset, Fig.~\ref{fig4}b). 

We examine two mechanisms for water passage through the pore. As with single-file water in carbon nanotubes, water molecules can move through the pore in a translocation mechanism, crossing the membrane while maintaining an unbroken chain of H-bonds~\cite{berezhkovskii_single-file_2002,chou_how_1998,corry_designing_2008,suk_water_2010,cohen-tanugi_water_2012,kalra_osmotic_2003,hummer_water_2001,mukherjee_single-file_2010,casieri_evidence_2010}. In the case of an atomically thin channel, however, water molecules can also cross the sheet individually, severing H-bonds and moving through the pore in an evaporation--condensation mechanism.  

To differentiate these mechanisms, we focus on the H-bond between two water molecules in the pore. The translocation mechanism relies on this \hb\ staying intact, while the evaporation--condensation mechanism requires that this bond breaks. We approximate the breaking rate of this \hb, \wb, as the Markov transition probability per unit time from a fully occupied pore to a singly occupied pore~(see the SI),
\begin{equation}\label{wbreak}
\wb \approx W_{\mathrm{full\rightarrow top}} + W_{\mathrm{full\rightarrow bottom}},
\end{equation}
using the state labels in Fig.~\ref{fig4}a. \wb\ follows the permeability closely~(Fig.~\ref{fig4}b); a larger \wb\ correlates to a higher permeability. This implies that the evaporation-condensation mechanism becomes more prevalent at higher permeability~(positive voltages). The steady-state occupancies of the Markov process support this picture as well: the probability of observing a singly occupied pore correlates positively with the permeability, while the probability of a observing a doubly occupied pore is anticorrelated with it~(Fig.~\ref{fig4}c). 

We propose the following picture to explain the results of the Markov model: When graphene is negatively charged~(positive voltage), it functions as an {\hb} acceptor and can form contacts with the positively charged hydrogens on the water molecules~(inset, Fig.~\ref{fig4}c). With their {\hb}s satisfied through contacts on the sheet, the water molecules can break their {\hb}s with other water molecules more easily. A positive voltage thus facilitates {\hb} breakage both between the water molecules in the channel and between the bulk and the channel waters, thereby lowering the barrier for the evaporation--condensation mechanism relative to the translocation one. Because water molecules pivot around a massive oxygen, there is an intrinsic molecular asymmetry in the dynamics of passage, so that the hydrogens enter the channel first.  We propose that the decrease in permeability at positive charge~(negative voltage) is due to an increasing energetic penalty for the light and rotationally mobile hydrogen atoms to enter the pore.

In this manuscript, we described a simulation method for atomistic systems under flow that is firmly rooted in constraint dynamics. In the low Re limit studied here, GD performs similarly when compared to the pump method and to linear response theory.  But from a practical perspective, simulations using GD consistently yield smaller standard errors for both permeabilities and effective viscosities when all other variables are the same~(see the SI). While the focus in this manuscript was on nanoscale permeability, it is not at all obvious that the three methods studied here will give similar results for other observables, particularly at high Re ($\mathrm{Re}>10$).  Indeed, GD always dissipates less heat than the pump method for the same mass flux.  This effect is likely due to heating at the discontinuity in the applied force used in  the pump method. These artifacts in the pump method may make GD more accurate at high Re~(see the SI) and for other observables more sensitive to heat flux.

With the appropriate methods in place, we studied the permeability of a nanopore embedded in a graphene sheet.  Permeability is not a simple function of the sheet's hydrophobicity. A Markov model reveals that the asymmetry of the permeability as a function of voltage can be explained in molecular terms, by a transition from a concerted translocation transport mechanism to an evaporation--condensation mechanism. Because the transport process is bottlenecked by only a few water molecules for pores of these sizes, the collective aspects of hydrophobicity have little bearing on the dynamics of water passage.

\begin{acknowledgements}
The authors thank Joe Ostrowski for preliminary work on the GD method. All simulations were done with the LAMMPS package~\cite{plimpton_fast_1995}, which we modified to perform GD\@. Simulation snapshots were made with the VMD and Tachyon packages~\cite{humphrey_vmd_1996,stone_efficient_1998}. This work utilized the Janus supercomputer, which is supported by the National Science Foundation~(Award Number CNS-0821794) and the University of Colorado Boulder. The Janus supercomputer is a joint effort of the University of Colorado Boulder, the University of Colorado Denver, and the National Center for Atmospheric Research. This material is based upon work supported by the National Science Foundation under Grant No.\ 1455365 and a Graduate Research Fellowship under Grant No.\ DGE-1144083. The authors declare no competing financial interests.
\end{acknowledgements}

\bibliography{paper}

\end{document}